\documentclass[prb,twocolumn,superscriptaddress]{revtex4-1}
\usepackage{graphicx}
\usepackage{dcolumn}
\usepackage{bm}
\usepackage{color}
\usepackage{amsmath, amsthm, amssymb,stmaryrd}
\usepackage{polski}
\usepackage[polish,french,english]{babel}
\newcommand{\rto}[1]{$R_2$Ti$_2$O$_7${#1}}
\newcommand{\tto}[1]{Tb$_2$Ti$_2$O$_7${#1}}
\newcommand{\hto}[1]{Ho$_2$Ti$_2$O$_7${#1}}
\newcommand{\dto}[1]{Dy$_2$Ti$_2$O$_7${#1}}

\usepackage{amssymb}
\usepackage{epstopdf}
\newcommand{\ra}[1]{\renewcommand{\arraystretch}{#1}}

\begin{document}

\title{Neutron Larmor diffraction investigation of the rare earth pyrochlores \rto{} ($R$ = Tb, Dy, Ho)}
\author{M Ruminy}
\email{martin.ruminy@gmx.de}
\affiliation{Laboratory for Neutron Scattering and Imaging, Paul Scherrer Institut, 5232 Villigen PSI, Switzerland}
\author{F Groitl}
\affiliation{Laboratory for Neutron Scattering and Imaging, Paul Scherrer Institut, 5232 Villigen PSI, Switzerland}
\affiliation{Laboratory for Quantum Magnetism, \'Ecole Polytechnique F\'ed\'erale de Lausanne (EPFL), CH-1015, Switzerland.}
\author{T Keller}
\affiliation{Max-Planck-Institut f\"ur Festk\"orperphysik, 70569 Stuttgart, Germany}
\affiliation{Max Planck Society Outstation at the FRM II, 85748 Garching, Germany}
\author{T Fennell}
\email{tom.fennell@psi.ch}
\affiliation{Laboratory for Neutron Scattering and Imaging, Paul Scherrer Institut, 5232 Villigen PSI, Switzerland}

\date{\today}

\begin{abstract}
In this work we present a neutron Larmor diffraction study of the rare earth pyrochlores \rto{}, with $R$ = Tb, Dy, Ho.  We measured the temperature dependence of the lattice parameter with precision $10^{-5}$, between 0.5 K and 300 K in each of the three compounds.  The lattice parameter of the spin ices \dto{} and \hto{} enters into the derivation of the charge of the emergent magnetic monopole excitations suggested to exist in these materials.  We found that throughout the range of applicability of the theory of emergent monopoles in the spin ices there will be no renormalization of the monopole charge due to lattice contraction.  In \tto{} strong magnetoelastic interactions have been reported.  We found no sign of the previously reported negative thermal expansion, but did observe anomalies in the thermal expansion that can be correlated with previously observed interactions between phonon and crystal field excitations.  Other features in the thermal expansion of all three compounds can be related to previously observed anomalies the elastic constants, and explained by the phonon band structure of the rare earth titanates.  The temperature dependence of the lattice strain in all three compounds can be correlated with the thermal population of excited crystal field levels.  

\end{abstract}

\maketitle

\section{\label{sec:Introduction}Introduction}

In dipolar spin ices such as \dto{} and \hto{}, the interplay of the local $\langle 111\rangle$ anisotropy with the dipolar and superexchange interactions is such that the groundstate of a single tetrahedron obeys the ice rule (two spins point in and two point out)~\cite{Bramwell:2001tpa}.  The ice rule is a local constraint which can be coarse grained to a non-divergent field~\cite{Huse:2003kh,Isakov:2004iu,Henley:2005iw} - an example of a Coulomb phase~\cite{Henley:2010vo,Fennell:2009ig} - and the excitations, which are local violations of this rule, can be identified as emergent magnetic monopoles~\cite{Castelnovo:2008hb}.  The monopolar nature of the excitations and their charge is best exposed by the so-called dumbbell model~\cite{Castelnovo:2008hb}.  Each magnetic moment is replaced by a dumbbell carrying a magnetic charge $\pm q$ at each end.  Satisfaction of the ice rule corresponds to ensuring charge neutrality at the center of each tetrahedron, while flipping a spin produces an equal and opposite excess of charge at two adjacent tetrahedron centers.  Subsequent  spin flips can restore the ice rule at the center of tetrahedra, in the wake of a hopping monopole excitation, so that low temperature properties of a spin ice can be described by a magnetic Coulomb gas.   The magnitude of the charge $q$, which is an essential property in theories of Coulomb gases, can be derived from the size of the magnetic moment $\mu$, and the separation of the tetrahedron centers, which is expressed in terms of the lattice constant of the diamond lattice formed by the tetrahedron centers, $a_d$: $q=\pm\mu/a_d$ (where $a_d=\sqrt{3/2}a$)~\cite{Castelnovo:2008hb}.  If the lattice constant (or magnetic moment) were to depend on temperature within the window of the effective theory, this could introduce a renormalization of the monopole charge which should be quantified.  However, to the best of our knowledge, the low temperature lattice parameters have not been accurately measured in \dto{} and \hto{}.

\tto{} is closely related to the spin ices, but has a quite contrasting low temperature behavior, in which a correlated but disordered magnetic state remains fluctuating to the lowest measured temperatures~\cite{Gardner:2010fu}.  It is not understood how this frustration can occur, since the antiferromagnetically interacting $\langle 111\rangle$ Ising moments of \tto{} should realize the unfrustrated counterpoint of the ice rule (all four spins point either in or out of the tetrahedron).  One aspect of the low temperature state which has been emphasized in some experimental works, and which is not a feature of the spin ices, is the strong magnetoelastic coupling.  In the temperature window where the spin correlations build up (i.e. $T\lesssim30$ K)~\cite{Fennell:2012ci}, various elastic properties~\cite{MAMSUROVA:1988wg,Nakanishi:2011bz} also become anomalous, and spin and lattice excitations are hybridized~\cite{Fennell:2014gf}.  Surprisingly, this magnetoelastic coupling apparently does not resolve the frustration.  One particular demonstration of this coupling was the measurement of developing lattice strain using single crystal x-ray diffraction~\cite{Ruff:2007hf}.  Again in the temperature window where the spin correlations develop, certain structural Bragg peaks were found to broaden anisotropically.  In the same experiments, it was observed that \tto{} enters a regime of negative thermal expansion at low temperature ($T\lesssim20$ K)).  This latter result has been debated, with x-ray powder diffraction experiments employed to show that there is no negative thermal expansion~\cite{Goto:2012br,Reotier:2014dn}.  Here we use the neutron Larmor diffraction technique, which provides access to both the thermal expansion and lattice strain in the same experiment, to revisit this question.

In the following, we briefly outline the neutron Larmor diffraction technique for interested readers; summarize our experimental methods; before presenting the results of our experiments on the temperature dependence of the lattice parameter and strain of \tto{}, \dto{}, and \hto{}; followed by discussions and conclusions.

\section{Larmor Diffraction}

\begin{figure}
\includegraphics[scale=1]{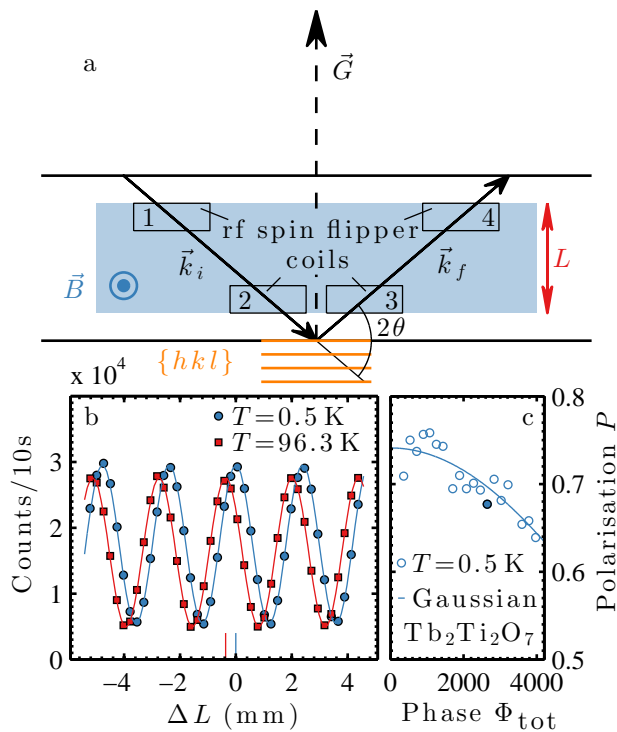}
\caption{Larmor diffraction setup and measurement. Panel a: Sketch of the Larmor diffraction technique. The basic principle is described in the main text. Panel b: Shift of the Larmor phase due to thermal expansion of the lattice of \tto{} measured by scanning the position $\Delta L$ of the final radio-frequency (rf) spin flipper coil at fixed frequency $f=200$\,kHz. Here, the thermal expansion of the lattice of \tto{} amounts to $3.66\times 10^{-4}$ between $T=0.5$\, and 96.3\,K. Panel c: Phase dependence of the polarization in \tto{} at $T=0.5$\,K. The resulting width of the lattice spacing distributions is $3.13\times10^{-4}$, and the filled point indicates the phase (frequency) at which the precession scans presented in Panel b were performed. }
\label{fig:larmor_raw}
\end{figure}

This section provides a brief introduction to neutron Larmor diffraction (LD) for readers who are unfamiliar with the technique, full details can be found in Refs.~\cite{Rekveldt:1999ft,Rekveldt:2001bt,Repper:2010ue,Repper:iPIOjmDe}.  The main purpose of LD is to obtain precise values of the lattice constant, or relative changes of it (i.e. $\Delta d/d$), and it also provides the distribution of $d$-spacings (which we call $\sigma_d$).  Readers not concerned with the details of how these are obtained can just look at the measurements of these quantities reported in section~\ref{sec:Results}.

Fig.~\ref{fig:larmor_raw} summarizes the basic principle of the LD experiment, and shows examples of the typical types of data which are obtained from a LD experiment.   The incident neutron beam is polarized, and the polarization of the scattered beam is analyzed.  The sample is held in the conventional condition  for neutron Bragg diffraction (i.e. the scattering vector $\vec{G}$ bisects $\vec{k}_i$ and $\vec{k}_f$), but before and after the sample the spin of the neutrons is made to precess, as the neutrons traverse regions of precisely known dimensions (further details follow).    In contrast to the neutron spin echo technique, however, the precession regions are such that total the Larmor phase $\Phi_{\rm tot}$ adds up. Since $\Phi_\mathrm{tot}$ depends on the path length, and therefore the $d$-spacing, as $\Phi_\mathrm{tot}=2m\omega_LLd/\pi\hbar$, and has a typical magnitude of 10$^3$-10$^4$ rad, LD is an elegant method to probe small changes in the lattice parameter.  Limitations in conventional diffraction measurements of the lattice spacing, such as beam divergence, are overcome by the Larmor labelling, and the technique can be used throughout the extremes of the parameter space accessible by neutron scattering, for example down to low temperatures or as a function of pressure \cite{Pfleiderer:2007hw}.

The precession region is actually defined by four radio-frequency spin flipper coils, as in neutron resonance spin echo~\cite{Gahler:1987gj}.  The coils are oriented precisely parallel to the lattice planes, which has the practical advantage of creating sharp boundaries for the precession region.  Three quantities can be varied in the experiment: an external parameter (in our case the temperature), the precession frequency $\omega_L$ (set by the rf flipper frequency), and the path length $L$ (modified by driving the final flipper coil backwards or forwards along the spectrometer arm).  Modification of $L$ at fixed $\omega_L$ effectively controls the precession time of the neutrons, such that when the final polarization is analyzed, a sinusoidal intensity dependence is obtained as the coil position scans the phase of the precession.  An example of this type of measurement is shown in Fig.~\ref{fig:larmor_raw}b.  The phase shift is due to a change in the lattice spacing, which we can extract from the relation $\Delta\Phi=\Phi_\mathrm{tot}\Delta d/d$.   

Measuring the polarization as a function of the total Larmor phase $\Phi_{\mathrm{tot}}$ yields additional information on the crystal lattice.  $\Phi_\mathrm{tot}$ depends on the Larmor frequency, which is set by the rf-flipper frequency.  For a given physical configuration of the instrument and sample, the polarization decays at higher frequencies due to dephasing by the distribution of $d$-spacings in the sample.  In fact, the polarization as a function of $\Phi_\mathrm{tot}$ is the cosine Fourier transform of the lattice spacing distribution $\sigma_d$. An example for this measurement is presented in Fig.~\ref{fig:larmor_raw}c. Generally, both the frequency dependence of the polarization and $\sigma_d$ are assumed to be Gaussian, hence their widths are connected inversely.

\section{\label{sec:Experimental_methods}Experimental methods}

The rare earth titanate pyrochlore samples used for this experiment have all been previously reported.  The crystal of \hto{} is a large single crystal ($\simeq 7$ g) grown in an image furnace and post-annealed under oxygen~\cite{Prabhakaran:2011hl}.  It was used in investigations of the diffuse scattering~\cite{Fennell:2009ig}.  The crystal of \tto{} is also large ($\simeq 7$ g), and has been used in investigations of diffuse scattering~\cite{Fennell:2012ci} and magnetoelastic excitations~\cite{Fennell:2014gf}.  Its characterization by x-ray diffraction and specific heat measurements have been discussed in Refs.~[\onlinecite{Fennell:2014gf}] and~[\onlinecite{mem_all_samps}] (it does not have the specific heat peak found in some \tto{} samples~\cite{Taniguchi:2013fi,Taniguchi:2015}).  The \dto{} crystal~\cite{Balakrishnan:1999er} is somewhat smaller ($\simeq 1.5$ g) because it is isotopically enriched with $^{162}$Dy to reduce the large absorption cross section of natural isotopic abundance dysprosium. It is the same crystal used for the measurement of magnetic diffuse scattering~\cite{Fennell:732176}, but before the LD study it was re-annealed in oxygen to eliminate possible oxygen vacancies~\cite{Sala:2014kz}. All samples were mounted on ultra-pure copper mounts to ensure good equilibration.  The crystals were aligned with the $[1\bar{1}0]$ axis vertical, so that the scattering plane contains $(h,h,l)$ wavevectors.

The LD measurements were performed at the thermal neutron resonance spin-echo spectrometer TRISP~\cite{Keller:2002bd,Keller:2007cv} at FRM II. Samples were loaded in a $^3$He sorption system with a base temperature of 0.5 K, with helium exchange gas added above $T = 5$ K to extend the temperature range for data collection up to room temperature.   For \tto{}, an incident wave vector of $k_i = 3.029$ \AA$^{-1}$ was used to perform LD measurements at the $(0,0,8)$ Bragg reflection. The same Bragg reflection was measured in the spin ices, but to account for their slightly smaller lattice constants, the incident wave vector  was adjusted to $k_i = 3.036$ \AA$^{-1}$.  This method allowed us to maintain a constant  scattering angle of  $2\theta=-110^\mathrm{o}$ and therefore constant tilting angles of the precession coils throughout the experiment - it is desirable to maintain the same scattering geometry for all samples in order to minimize sources of error.  A perfect germanium single crystal with a well-determined lattice constant of $a = 5.65728(7)$ \AA~ was used as a reference sample (again adjusting $k_i$ to measure the $(0,0,4)$ reflection without repositioning the spectrometer).

The total phase $\Phi_{\mathrm{tot}}$ was measured as a function of temperature, by scanning the position of the last precession coil to vary the length $L$ of the effective precession region.  From the change of the phase relative to the lowest temperature reference, we obtain the relative thermal expansion $\Delta d/d$.  The thermal expansion was measured in a single temperature sweep, maintaining a fixed flipper frequency of 200 Hz, in order to ensure the stability of the radio-frequency and temperature of the coils.  Before or after stabilization of the set up for a temperature sweep, we visited selected points across the full temperature region, where we determined the distribution of the lattice spacings $\sigma_d$.  These were obtained from measurements of the final polarization as function of the spin flipper frequency and hence the total Larmor phase $\Phi_\mathrm{tot}$.

\section{\label{sec:Results}Results}

\subsection{Thermal expansion}

\begin{figure}
\includegraphics[width=0.4\textwidth]{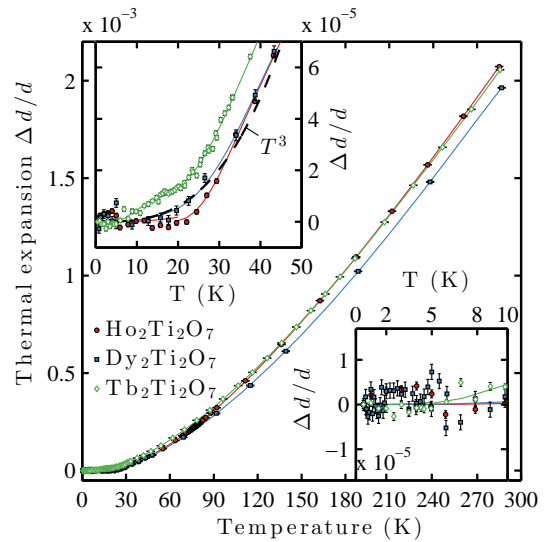}
\caption{Thermal expansion of \rto{}, with $R=$Ho, Dy and Tb as extracted from the measurement of the Larmor phase shift. The data is presented relative to the corresponding lattice size at $T=0.5$\,K.  The solid lines are interpolations used in the construction of the thermal expansion coefficient $\alpha$, as described in the text, but it can be seen that in \dto{} below 50 K, the expansion is well described by a cubic temperature law (emphasized in the upper inset, where only every fourth data point is shown for \dto{}).  While \hto{} is similar, \tto{} shows an anomaly below 20\,K. At temperatures below 10\,K (lower inset), the thermal expansion is zero within the uncertainty bandwidth of $\Delta(\Delta d/d)=10^{-5}$.}
\label{fig:larmor_tdep}
\end{figure}

\begin{figure}
\includegraphics[width=0.4\textwidth]{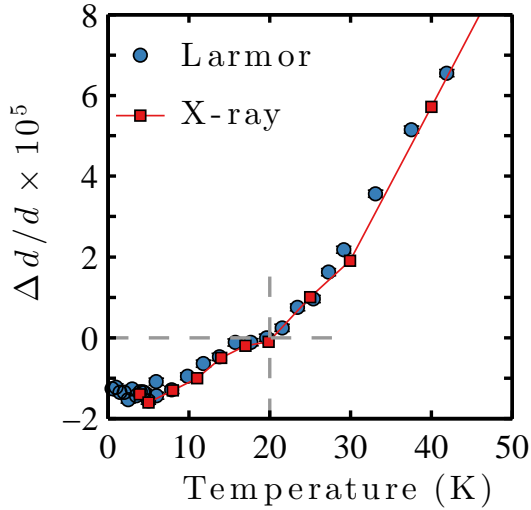}
\caption{Comparison of the thermal expansion of a single crystal of \tto{}, extracted from these LD measurements, with that obtained by powder x-ray diffraction~\cite{Goto:2012br}.  The two data sets are normalized to one another at $T=20$ K.}
\label{fig:larmor_xray}
\end{figure}

The temperature dependence of $\Delta d/d$ for all three compounds is shown in Fig.~\ref{fig:larmor_tdep}.  For $0.5<T\lesssim5$ K, no thermal expansion can be distinguished in any of the compounds, within the resolution of Larmor diffraction (which is in principle of the order of $10^{-6}$, but here, and generally in practice, is of the order of $10^{-5}$).  The error on the lattice parameter at a particular temperature is kept small by the robust fitting of the sinusoidal intensity function over four periods.  When we report the absolute lattice parameters (see below), the error bar originates from the spread of repeated observations visible in the second inset of Fig.~\ref{fig:larmor_tdep}.  

In the temperature region $5<T\lesssim40$ K, the thermal expansion of \dto{} follows a $T^3$ dependence, as might be expected for a system whose lattice and thermal properties are purely controlled by acoustic phonons in this temperature range.  There is no obvious reason why this is not also the case for \hto{}. Previously, in \tto{}, negative thermal expansion was reported in the temperature range  $0.3<T<20$ K~\cite{Ruff:2007hf}.  This observation was made using single crystal x-ray diffraction, and has subsequently been disputed, based on the results of x-ray powder diffraction experiments~\cite{Goto:2012br,Reotier:2014dn}, which do not show the effect.  Likewise, our data also do not show any evidence of negative thermal expansion, but they do clearly show an anomaly in the temperature dependence at $T\approx 22.5$ K.  Although the lattice of \tto{} begins to expand at $T\approx 5$ K, the expansion is interrupted by a plateau at $T\approx  22.5$ K, before expansion continues as the temperature is increased further.  The feature cannot easily be discerned in the powder diffraction data as it is presented in Ref.~[\onlinecite{Goto:2012br}], but in a comparison on the same scale it is clearly present, as shown in Fig.~\ref{fig:larmor_xray}.  This feature is not discussed in Ref.~[\onlinecite{Goto:2012br}], while it is qualitatively described, but not shown, in Ref.~[\onlinecite{Reotier:2014dn}].  

The lattices of all three compounds continue to expand smoothly at higher temperatures (i.e. $50<T<300$ K).  It is notable that the relative expansion of \hto{} accelerates more rapidly than either \tto{} or \dto{}, such that while it is the smallest at $T=20$ K, it has become the  largest at $T=270$ K. When looking at the curves of $(\Delta d/d)$ vs. $T$ in Fig.~\ref{fig:larmor_tdep}, one may gain the impression that the rates of expansion of the three compounds have subtle departures from monotonicity.  For example, between 30 and 150 K, the curves for \tto{} and \hto{} seem to approach one another, then separate, and then converge and cross.  Since the three compounds have identical structures and closely related phonon band structures~\cite{phonons}, the simple expectation would be that they expand identically, with some small renormalization related to the change in mass of the rare earth ion.  

\begin{figure}
\includegraphics[width=0.4\textwidth]{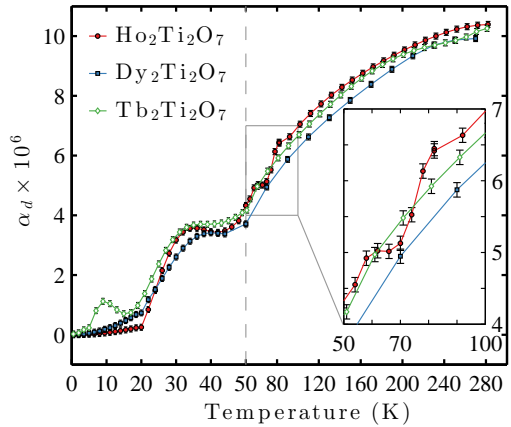}
\caption{Thermal expansion coefficient of \rto{}, with $R$ = Tb, Dy, and Ho, as extracted from the temperature dependence of $(\Delta d/d)$.}
\label{fig:larmor_alpha}
\end{figure}

To try to clarify this point, we computed the derivative of the thermal expansion curves and present $\alpha_d(T)=(1/d)(\delta d/\delta T)$ in Fig.~\ref{fig:larmor_alpha}.  The thermal expansion data of Fig.~\ref{fig:larmor_tdep} was piecewise interpolated with weighted polynomials.  The derivative of the interpolated thermal expansion is calculated with a centered five-point stencil, and three-point or one-sided numerical derivatives at its ends.  The derivatives can depend on the spacing of the temperature points ($dT)$ used in the interpolation, and the degree of the polynomial.    We have tested the effect of both parameters to establish sensible values~\footnote{For $T>50$ K the data is best described by fixing the polynomial order to 6, and using a large temperature spacing ($dT=10$ K for $R=$Ho,Tb, and $dT=20$ K for $R=$Dy).  The inflection point at $T\approx 67$ K for $R=$Ho can be washed out when using large $dT$ but is otherwise a robust feature, which becomes more (less) pronounced when decreasing (increasing) $dT$, here we used $dT=4$ K, and polynomial order $> 6$.  The interpolation of $\delta_d/d$ for $T<50$ K requires a polynomial order $> 6$, in \tto{} the entire temperature range $5<T<50$ K is described by one polynomial, while in the spin ices only the range $20<T<50$ K,  in both cases $dT = 2$ K.}.

\begin{figure*}
\includegraphics[scale=1]{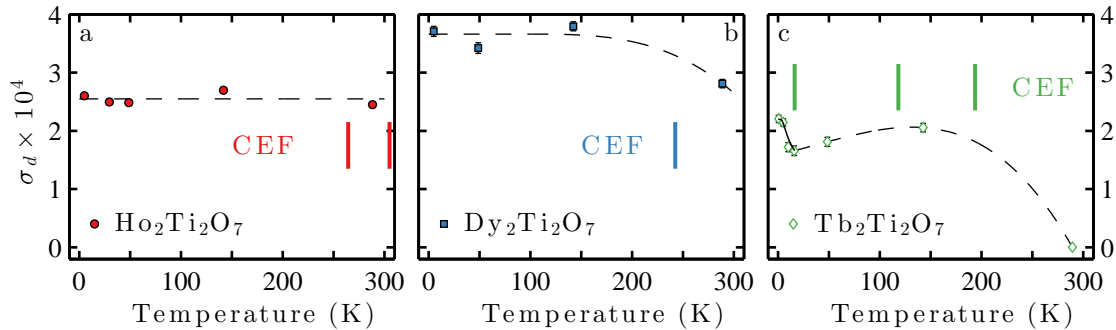}
\caption{Temperature dependence of the full width at half maximum of the lattice spacing distributions $\sigma_{d}$ measured in \rto{}, with $R=$ Tb, Dy and Ho (a,b and c). The vertical lines denote the temperatures of CEF excitations in the respective compound, while the broken lines are guides to the eye. All curves are normalized to the $T=300$\,K measurement of \tto{}, for which $P$ was maximal. The distinct features in the temperature dependence of $\sigma_{d}$ in \tto{} are interpreted as signatures of magnetoelastic coupling, as described in the main text.}
\label{fig:larmor_strain}
\end{figure*}

Given this treatment, features in the temperature dependence of $\alpha(T)$ should be interpreted carefully, but the features we discuss appear to be reasonably robust, and this presentation helps to emphasize effects which are not immediately visible in $(\Delta d/d)(T)$.   Foremost, for $T<20$ K, we see a clear difference between \tto{}, which has a pronounced maximum, and the spin ices, which do not.   A broad maximum at $T\approx 35$ K is clearly visible in all three compounds.  For $T>50$ K, and above, the comparison is more complicated.  In our measurements of \hto{}, there is a particularly high density of points around $T\approx80$ K.  An anomaly in the elastic constants of all three compounds was reported at $T\approx80$ K~\cite{Nakanishi:2011bz}.  In \hto{}, where the large crystal and good count rate enabled many measurements, we searched for a sign of it in the thermal expansion.  Nothing dramatic can be seen in the temperature dependence of $(\Delta d/d)(T)$ in Fig.~\ref{fig:larmor_tdep}, but when we consider the derivative, we see clear inflections in $\alpha_d(T)$, as shown in the inset of Fig.~\ref{fig:larmor_alpha}.  However, for \tto{} and \dto{}, we have far fewer data points in this temperature range, so the absence of this feature in $\alpha$ for these compounds cannot be regarded as definitive. There is a minimum in $\alpha$ for \tto{} at $T\approx 240$ K, but not in the spin ices.  We will discuss the origin of all these features below.

\subsection{Lattice strain}

Another quantity which can be obtained from Larmor diffraction is the $d$-spacing distribution, $\sigma_d$, or lattice strain.  In Fig.~\ref{fig:larmor_strain}, we show how $\sigma_d$ depends on temperature in each of the three compounds.  The lattice strain of \tto{} at 297 K was found to be the least, i.e. the decay of the neutron polarization with increasing precession frequency (the quantity shown in Fig.~\ref{fig:larmor_raw}c) is least.  All data is normalized to this point, which therefore appears at zero.  In \tto{}, $\sigma_d$ is not monotonic with increasing temperature.  A marked drop in $\sigma_d$ occurs at $T\approx 20$ K, but it then seems to pass through a subsidiary maximum at $T\approx150$ K, and drops significantly by $T \approx 300$ K.  In \dto{}, there is a clear decrease in $\sigma_d$ at the highest temperatures, while in \hto{}, little or no effect is visible.  The energy scales of the crystal field excitations of each compound~\cite{xtal_fields} which fall in the temperature window of the experiment are also shown in Fig.~\ref{fig:larmor_strain}, and we will discuss their role below.  

\subsection{Absolute lattice parameters}

\begin{figure}
\includegraphics[width=0.4\textwidth]{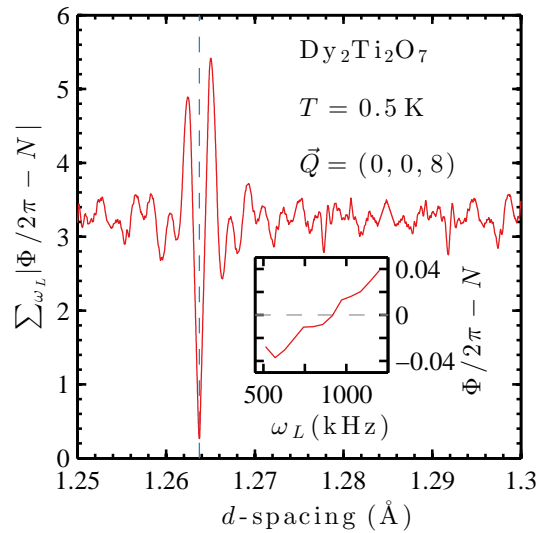}
\caption{Example of the determination of the absolute lattice parameter of \dto{}.  The best fitting integer precession number ($N$) indicates the value of the $d$-spacing.  The inset displays the (frequency-dependent) summands at the minimum of the sum displayed in the main figure. It is a verification that we found the correct d-spacing, because at this value of $d$, the total phase becomes an integer for all frequencies simultaneously.}
\label{fig:larmor_absolute}
\end{figure}

An important aim of this experiment was to obtain the absolute value of the lattice constant in the spin ices in order to quantify its effect on the emergent magnetic monopole charge.  In order to determine absolute values of the lattice constant from Larmor diffraction, an accurate reference is required, in this case the germanium crystal mentioned above.  When the scattered polarization is maximal, neutrons undergoing Bragg scattering in the crystal have precessed in total an integer number of times while traversing the precession fields of length $L$.  In consequence, the total Larmor phase $\Phi_{\rm tot}$ for a given flipper frequency and corresponding $L$ becomes an integer only if $d$ is the absolute lattice spacing. The value of the absolute $d$-spacing is then found by minimizing the value of the total phase versus its nearest integer in the region of the approximately known $d$-spacing. The length $L$ of the precession field, which is required with high precision, was measured using the above mentioned germanium reference sample and applying the relations in reverse.  An example of this procedure is shown in Fig.~\ref{fig:larmor_absolute} for \dto{}, and the relevant values of the lattice parameters are given in Table \ref{tab:absolute_a}.  We have also recently determined the room temperature lattice parameters of each compound by synchrotron powder x-ray diffraction with calibration against a NIST silicon standard~\cite{phonons,mem_all_samps}.  Those lattice parameters are also shown in the table, and agree approximately, i.e. at the level of $10^{-3}$, with those determined by Larmor diffraction (N.B. for \hto{} and \dto{} the x-ray values are for powder samples, while the value for \tto{}, which has a sample dependent lattice parameter, is from a fragment of the same crystal (which is known as MH1 in Ref.~[\onlinecite{mem_all_samps}])).

\begin{table} 
 \caption[Absolute lattice constants $a$ of \rto{} in \AA{} determined from neutron Larmor diffraction at different temperatures.]{
Absolute lattice constants $a$ of \rto{} in \AA{} determined from neutron Larmor diffraction at different temperatures.  The precision of the lattice parameters is $\approx \pm5\times10^{-6}$.\label{tab:absolute_a}}
\centering
\ra{1.3}
\begin{tabular}{ r r r r }\hline\hline
     \multicolumn{1}{r}{$T$\,(K)} & \multicolumn{1}{r}{\hto{}} & \multicolumn{1}{r}{\dto{}} & \multicolumn{1}{r}{\tto{}}\\  \hline
  0.5 & $10.08344$ & $10.10984$ & $10.13032$  \\
  48 & $10.08424$ & $10.11048$ & $10.13152$  \\
  142 & $10.09040$ & $10.11632$ & $10.13792$  \\ 
  289 & $10.10464$ & $10.13000$ & $10.15192$  \\ \hline
  297.15 (x-ray) & 10.102 & 10.130&10.155 \\
  \hline \hline
\end{tabular}
      \end{table}

\section{\label{sec:Discussion}Discussion}

Our central aim, to understand the effect of possible lattice expansion on the monopole charge in spin ice, was realized - no relative thermal expansion occurs in \dto{} or \hto{} below 5 K.  This more or less coincides with the expected window of applicability of the monopole theory, so no charge renormalization needs to be incorporated.  (At higher temperatures the increasing population of double charge monopoles, and the decreasing validity of the Ising approximation mean that the effective theory becomes progressively less applicable.)  We advance our values of the lattice parameter at 0.5 K, as shown in Table~\ref{tab:absolute_a}, as the most appropriate value to use in calculating the monopole charge.  (We recently advanced appropriate values of the magnetic moment derived from the crystal field ground state wavefunctions in Ref.~\onlinecite{xtal_fields}.) 

We did not observe any negative thermal expansion in \tto{}, but we did observe an anomaly in the lattice contraction.  Previously, we have shown that the first excited crystal field (CEF1 at 1.5 meV or 17.4 K) is coupled with a transverse acoustic phonon forming a hybrid magnetoelastic mode (MEM), and this coupling develops at $10<T\lesssim25$ K.  The plateau in $(\Delta d/d)$ (or minimum in $\alpha$) occurs exactly in the temperature range where the hybridization develops~\cite{Fennell:2014gf,mem_all_samps}, and the appearance of this feature only in \tto{} is a clear sign that it originates in a spin-lattice effect that is absent in the other two compounds.   An important point about this observation is that it was made in a single crystal of \tto{}.  A central criticism of the observation of negative thermal expansion~\cite{Ruff:2007hf,Goto:2012br,Reotier:2014dn} was that the properties of \tto{} single crystals are variable, while powders are reproducible, though none of the samples involved were characterized to the degree that \tto{} samples typically are now~\cite{Taniguchi:2013fi,Taniguchi:2015,Wakita:2015vr,Fennell:2014gf,Guitteny:2015vo,Kermarrec:2015gz,mem_all_samps}.  The appearance of the expansion anomaly in two generic \tto{} powders, and a generic single crystal, along with the consistent form of the excitation spectrum in different \tto{} crystals~\cite{mem_all_samps}, strongly suggest that the magnetoelastic phenomena observed in all \tto{} samples in this temperature range are ubiquitous, and that the various low temperature behaviors of \tto{} all stem from the same magnetoelastic degrees of freedom~\cite{Fennell:2014gf} formed in the range $10<T<30$ K. 

The lattice strain reported by Ruff~{\it et al.}~\cite{Ruff:2007hf} seems to be a robust observation.  Since our experiment was conducted at the $(0,0,8)$ reflection, we are sensitive to the longitudinal broadening that they observed at $(0,0,12)$.  We observed a strong increase in the distribution $\sigma_d$ below 20 K, in agreement with the findings of Ruff~{\it et al.}.  Although the strain broadening has not been observed in the powder diffraction experiments, it is not obvious that it can be observed in a conventional powder diffraction experiment.   We see in Fig.~\ref{fig:larmor_strain} that the development of lattice strain at low temperatures in \tto{} occurs in the temperature window in which the MEM appears.  We also recently showed that the transition at 17 meV, or 197 K, is also coupled to a transverse optical phonon (TOP), forming a bound state~\cite{xtal_fields}.  Although we do not have a detailed temperature dependence of $\sigma_d$, it is clear that the lattice strain relaxes further at this temperature scale.  The temperature dependence of $\alpha_d$ is more detailed and does reveal a minimum at $T\approx240$ K, close to the energy of the crystal field/TOP bound state.  This feature also appears only in \tto{} and not the spin ices, indicating that it also originates in a spin-lattice coupling only present in \tto{}.  The minimum appears at the onset of the formation of the bound state, in close similarity with the minimum at $T\approx20$ K, at the onset of the formation of the MEM. Due to the much higher temperature of the bound state, its onset is spread over a larger temperature range than the one of the MEM, and hence the minimum is broader.

The relation between lattice strain and the coupling of excited crystal field levels with phonons has been investigated in compounds such as TmPO$_4$ and TmVO$_4$~\cite{MEHRAN:1976wl,MEHRAN:1977ts,MEHRAN:1982tn,MEHRAN:1983wv,Gehring:1975uk}.  It was found that if the phonons are coupled to degenerate (doublet) ground states, a cooperative Jahn-Teller transition occurs, while if the phonons are coupled to a degenerate excited state above a singlet ground state, no transition occurs since the ground state is any way unique, but lattice strains build up as the crystal field state is depopulated.  This points to a general explanation of the development of the lattice strains that we observed.  In \hto{}, the crystal field states are simply too high in energy to have any strong effect within the temperature window of this study.  In \dto{} the first excitation is a little lower, apparently sufficient to see a relaxation in the strain only at the highest measured temperature, while in \tto{}, the consequences of depopulating two excited states with important coupling to phonons can be seen.  It is interesting to note that the lattice strain is of similar magnitudes in all three compounds.  While it is now suggested that \tto{} crystals may be inhomogeneous~\cite{Wakita:2015vr}, we find that the lattice strain of our \tto{} crystal (and LD probes the bulk of a large crystal) is not generally larger than the other two compounds.  

In addition to the effects visible in $(\Delta d/d)$ and $\alpha_d$ which are unique to the spin-lattice coupling in \tto{}, there are temperature scales that appear to be relevant to all three compounds, and therefore derive from pure lattice effects.  The first is $T\approx 30$ K, where a prominent maximum in $\alpha_d$ clearly occurs in all three compounds.  The second is $T\approx75$ K, where there are inflections in $\alpha_d$ for \hto{}, which we correlate with an anomaly in the elastic constants that was reported by Nakanishi {\it et al}.~\cite{Nakanishi:2011bz} for all three compounds and suggested to be due to a structural transition of unknown type.  We do not have sufficient data to conclude that the effect we observe is unique to \hto{}, so suppose it to be the same effect observed by Nakanishi {\it et al.}, and to appear in all three compounds.  

All of these changes in the rate of thermal expansion appear to derive from the thermal population of bundles of optical phonons.  We recently reported the phonon band structure of the three compounds~\cite{phonons}, which is almost identical, and contains several low lying optical phonons at $E\approx6$ meV and $E\approx9.5$ meV.  The thermal expansion coefficient may be written in terms of the mode Gr\"uneisen parameters and temperature derivative of the phonon mode occupation numbers~\cite{Zwanziger:2007bf}.  For the optical modes in question, the two temperature scales correspond to rapid growth in the latter quantity.  No further evidence for the  structural phase transition at $T\approx75$ K proposed in Ref.~\onlinecite{Nakanishi:2011bz} has been advanced, and we now believe that the observed anomaly in the elastic constants is explained by a strongly temperature dependent contribution to the thermal expansion coefficient.  

Mamsurova {\it et al}. observed peaks in the internal friction of \tto{}, at 40 K and 110 K~\cite{MAMSUROVA:1986wx}, more or less corresponding to the boundary of the two lattice features in $\alpha_d$.  They could observe these peaks, at the same temperatures, in all rare earth titanate pyrochlores, including Y$_2$Ti$_2$O$_7$.  They related these to dielectric anomalies that they had previously observed~\cite{Mamsurova:1985ug}, stating that these also occurred at the same temperatures, though in fact they are at 30 K and 80 K, in close correspondence with our anomalies in $\alpha_d$.  These dielectric anomalies were attributed to the formation of two types of dipoles in the sample  with different frequencies, probably due to the motion of anion vacancies or interstitial cations, though no explanation of the temperature scales was made~\cite{Mamsurova:1985ug}.  All the modes in the low-lying optical phonon bundles have $u$ character~\cite{phonons}, and hence are dipole active.  We therefore suggest that the population of these modes is an alternative and plausible explanation for the anomalies observed in Refs.~\onlinecite{Mamsurova:1985ug,MAMSUROVA:1986wx}.  Firstly, we would expect defect motions to occur at rather higher temperatures; secondly, our explanation is also consistent with their other tests (the anomalies disappear in other rare earth titanate crystal structures, as they must since the phonon spectrum is different; the anomalies exist only in titanates and not zirconates, and as discussed in Ref.~\onlinecite{phonons}, replacement of titanium by zirconium also modifies the phonon density of states).

Finally, we point out that this is the first use of Larmor diffraction to study the rare earth titanates.  Our study covered the temperature range $0.5<T<300$ K, as it was convenient to use the $^3$He refrigerator to study all three compounds.  Recently it has been suggested that there is a quadrupole ordering transition in \tto{} at a temperature of $T\approx 0.5$ K~\cite{Takatsu:2016im}, only observable in crystals of very high quality or powders.  Larmor diffraction could be used to study the transition in a small crystal with directly verified heat capacity - since the signal is proportional to Bragg intensity, even a relatively small sample can be used.  A subtle lattice anomaly related to a quadrupole ordering might be detected in this way.

\section{\label{sec:Conclusion}Conclusion}

We have used Larmor diffraction to investigate the thermal expansion, lattice strain, and absolute values of the lattice constants of \tto{}, \dto{}, and \hto{}, in the temperature range $0.5<T<300$ K.  We report the absolute values of the lattice constants at low temperature in the context of the charge of the emergent magnetic monopole excitations in spin ice.  We found no evidence of negative thermal expansion in \tto{}, but nonetheless observe various  anomalies in the thermal expansion $\Delta d/d$ and thermal expansion coefficient $\alpha_d$.  Some of these, in \tto{} could be related to interactions between crystal field and phonon excitations, while others that occur in all three compounds could be connected with the phonon band structure.  We did  confirm that lattice strain does increase at low temperature in \tto{}, as previously reported.  The general behavior of the lattice strain is also suggested to be related to the interaction of excited crystal field levels and phonons, if a compound has levels which may couple to phonons, in the temperature window of the experiment.

\acknowledgements{Neutron scattering experiments were carried out at the FRM-II, Munich, Germany.  Work at PSI was partly funded by the SNSF (Schweizerischer Nationalfonds zur F\"orderung der Wissenschaftlichen Forschung) (grant 200021\_140862 and 200020\_162626).}


%

\end{document}